%% file: main.tex
\documentclass[sigconf,nonacm]{acmart}

\setcopyright{none}

\usepackage{float}
\usepackage[most]{tcolorbox}
\usepackage{enumitem}
\usepackage{balance}

\input{preamble}

\begin{document}

\title{Tuning Qwen2.5-VL to Improve Its Web Interaction Skills}

\author{Alexandra Yakovleva}
\affiliation{%
  \institution{Aalto University}
  \city{Espoo}
  \country{Finland}
}
\email{alexandra.yakovleva@aalto.fi}

\author{Henrik Pärssinen}
\affiliation{%
  \institution{Aalto University}
  \city{Espoo}
  \country{Finland}
}
\email{henrik.parssinen@aalto.fi}

\author{Harri Valpola}
\affiliation{%
  \institution{System 2 AI}
  \city{Helsinki}
  \country{Finland}
}
\email{harri@system2ai.com}

\author{Juho Kannala}
\affiliation{%
  \institution{Aalto University}
  \city{Espoo}
  \country{Finland}
}
\affiliation{%
  \institution{University of Oulu}
  \city{Oulu}
  \country{Finland}
}
\email{juho.kannala@aalto.fi}

\author{Alexander Ilin}
\affiliation{%
  \institution{System 2 AI}
  \city{Helsinki}
  \country{Finland}
}
\email{alexilin@system2ai.com}

\hyphenation{Ale-xan-der}
\renewcommand{\shortauthors}{Alexandra Yakovleva, Henrik Pärssinen, Harri Valpola, Juho Kannala, \& Alexander Ilin}

\begin{abstract}
\input{abstract.tex}
\end{abstract}



\maketitle

\section{Introduction}
\label{sec:intro}
\input{intro.tex}

\section{Related works}
\label{sec:related_works}
\input{related_works.tex}

\section{Environment}
\label{sec:environment}
\input{environment.tex}

\section{Two-Stage Fine-tuning}
\label{sec:method_and_results}
\input{method.tex}

\section{Conclusion}
\label{sec:conclusion}
\input{conclusion.tex}

\balance

\bibliographystyle{ACM-Reference-Format}
\bibliography{references}

\end{document}

%% file: preamble.tex
\usepackage{xspace}
\usepackage{pifont}
\usepackage{standalone}
\usepackage{tikz}
\usetikzlibrary{positioning, backgrounds, calc, arrows.meta, fit}

\usepackage{multirow}

\makeatletter
\DeclareRobustCommand\onedot{\futurelet\@let@token\@onedot}
\def\@onedot{\ifx\@let@token.\else.\null\fi\xspace}

\def\eg{\emph{e.g}\onedot}

 \def\vs{\emph{vs}\onedot}

\makeatother

\newcommand{\secref}[1]{Section~\ref{#1}}

%% file: abstract.tex
Recent advances in vision–language models (VLMs) have sparked growing interest in using them to automate web tasks, yet their feasibility as independent agents that reason and act purely from visual input remains underexplored. We investigate this setting using Qwen2.5-VL-32B, one of the strongest open-source VLMs available, and focus on improving its reliability in web-based control. Through initial experimentation, we observe three key challenges: (i)~inaccurate localization of target elements, the cursor, and their relative positions, (ii)~sensitivity to instruction phrasing, and (iii)~an overoptimistic bias toward its own actions, often assuming they succeed rather than analyzing their actual outcomes. To address these issues, we fine-tune Qwen2.5-VL-32B for a basic web interaction task: moving the mouse and clicking on a page element described in natural language. Our training pipeline consists of two stages: (1)~teaching the model to determine whether the cursor already hovers over the target element or whether movement is required, and (2)~training it to execute a single command (a mouse move or a mouse click) at a time, verifying the resulting state of the environment before planning the next action. Evaluated on a custom benchmark of single-click web tasks, our approach increases success rates from 86\% to 94\% under the most challenging setting.

%% file: ccs_concepts.tex
%% file: intro.tex
Large language and multi-modal models have rapidly advanced in recent years, enabling the creation of capable web and GUI-based agents, yet most existing open-source systems remain either (i)~text-only agents that reason over HTML or DOM parsing or accessibility trees~\cite{deng2023mind2web,zhou2024webarena}, or (ii)~multi-module systems that decompose perception, planning, execution, and reflection on previous actions across specialized components \cite{he2024webvoyager,niu2024screenagent,bhathal2025websight}.
Text-only agents, although computationally efficient and widely studied, lack visual grounding and therefore struggle when textual descriptions do not accurately reflect the interface or when essential information is available only in an accompanying image or plot.
Multi-module designs show strong performance but introduce additional complexity and dependencies.
They often rely on large closed-source models, either as the core agent~\cite{zheng2024seeact,he2024webvoyager} or by outsourcing key modules such as planning or reasoning~\cite{bhathal2025websight}.

This study aims to investigate a more constrained but conceptually fundamental setting: \textit{how well a single open-source VLM can act as a web agent when relying solely on visual information?}
We build our agent on top of Qwen2.5-VL-32B \cite{qwen2.5-VL}, a mid-size model from the state-of-the-art open-source Qwen family. Our agent perceives browser screenshots and acts without access to the DOM or any external planner: analyzes the image, reasons about the task, and selects the next command, unifying perception, reasoning, and control within a single component.

Preliminary experiments with this model revealed several limitations.
First, it shows limited spatial grounding, often failing to determine whether the cursor hovers over the target and to predict accurate movement coordinates.
Second, it is sensitive to prompt phrasing, and when operating within a rich action space that includes scrolling, typing, and other commonly used operations, it may select inconsistent or unnecessary actions. 
Finally, the model tends to assume that its actions succeed without re-evaluating the screen, resulting in an overoptimistic bias driven by the interaction trace. 
This indicates open-loop behavior: the model often combines multiple atomic actions (\eg move+click+type) in a single step instead of following an action-by-action strategy that would allow self-correction.

These findings motivated a minimal yet representative browser environment for studying visual grounding and robustness to contextual noise.
We restrict the agent’s action space to two atomic operations: mouse move and mouse click, and focus on single-click tasks only. 
This setup forms a basis from which broader multi-step interactions can be extrapolated.

The contributions of this work are:
\begin{itemize}[leftmargin=*, topsep=2pt]
    \item A simple setup for studying vision-only web interaction using a custom collection of single-click tasks\footnote{\url{https://huggingface.co/datasets/alexandrayakovleva/single-click_bench}.}.
    \item A two-stage fine-tuning that improves visual grounding and addresses interaction-trace-induced bias and prompt dependence.
    \item An empirical evaluation demonstrating both the feasibility and the remaining challenges of purely visual web interaction.
\end{itemize}

%% file: related_works.tex
The first web and GUI agents based on large language models operated on text-only representations extracted from HTML, DOM, or accessibility trees for planning and acting~\cite{deng2023mind2web,zhou2024webarena}.
With the development of vision-language models, to improve grounding, later systems utilized visual information as an additional modality, combining screenshots with previously exploited page metadata~\cite{zhou2024webarena,zheng2024seeact}.
Several recent approaches explored purely visual control, but mostly through multi-component architectures that separate perception, reasoning, planning, and reflection~\cite{niu2024screenagent,bhathal2025websight,qin2025ui}, sometimes delegating key modules to closed-source models~\cite{bhathal2025websight}.

The closest to our setting is SeeClick~\cite{cheng2024seeclick}, which demonstrates that an open-source model can act from screen when additionally fine-tuned for visual grounding on various GUI and web page screenshots.
While this work shows that a single-module purely visual agent can succeed in computer-use tasks and highlights the importance of GUI grounding, we further analyze the behavior patterns, limitations, and failures of this agentic setup.

Our work provides a controlled and systematic study of this minimal vision-only web-agent setting, showcasing its key bottlenecks, such as spatial grounding, prompt dependence, and history-induced bias, and demonstrating how they can be mitigated through targeted fine-tuning.

%% file: environment.tex
\begin{figure}[tbp]
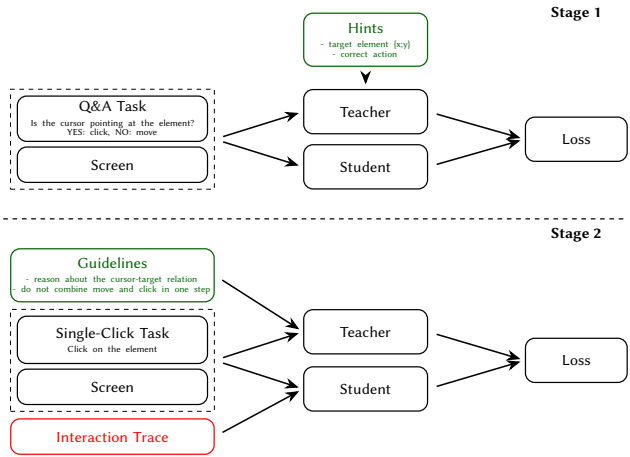

    \centering
    \includestandalone[width=\columnwidth]{fig/main_fig_tikz}    
    \caption{Overview of the proposed two-stage fine-tuning pipeline.
Stage 1 distills action selection from a teacher with access to privileged hints.
Stage 2 distills the desired interaction behavior for each step, including (i)~focusing on the current screen without being distracted by prior actions, (ii)~verifying whether the cursor already hovers over the target, and (iii)~selecting the next single action, move or click.}
    \label{fig:diagram}
\end{figure}

We built a custom simulation called VisionWebBrowser using Playwright in Python.
At each step, the agent receives (i) a textual task description, outlining the goal and available actions, (ii) a screenshot of the current webpage with a rendered macOS-style cursor, and (iii) optionally, a trace of its previous interactions -- reasoning and executed actions. 
Although unnecessary for single-click tasks, interaction history is essential in realistic multi-step settings. We therefore consider enabling it and address the performance degradation it causes.

The cursor rendering is what distinguishes our environment from existing ones, \eg, VisualWebArena~\cite{koh2024visualwebarena}.
Intuitively, the cursor serves as a visual anchor that helps the agent to determine its position relative to the target element and build causal relations between its actions and the resulting interface changes.

The VisionWebBrowser has two methods the agent can operate:

{\centering\texttt{mouse\_move}(x, y) \quad \text{and} \quad  \texttt{mouse\_click()}.\par}

\begin{samepage}
Interaction itself follows a ReAct-style loop~\cite{yao2023react}, where each step is implemented in two sub-steps:
\begin{enumerate}[leftmargin=*, topsep=2pt]
    \item \textbf{Reasoning sub-step.} The agent is prompted to produce its reasoning inside \texttt{<think>...</think>} tags.
    Additionally, we introduce behavioral guidance to focus on the current cursor-target relation: 
\textit{``(i) Describe the location of the target element.
\mbox{(ii) Locate} the cursor. Is it currently pointing at the target element?
\mbox{(iii) Decide} the next action based on their relative positions.
Cursor description: black arrow angled up-and-left with a thin white outline.''}

    \item \textbf{Action sub-step.} The agent is then prompted to output its action inside an \texttt{<ipython\_cell>...</ipython\_cell>} block.
    To keep outcomes analyzable, the agent is encouraged to call exactly one browser method per step: \texttt{mouse\_move} or \texttt{mouse\_click}.
\end{enumerate}
\end{samepage}
After the action is executed, the resulting webpage state, as well as the console output and error streams, is returned to the agent for the next iteration.
In~\secref{subsec:stage2}, we show that the behavioral guidance to analyze the cursor–target relation and separate move and click actions can be internalized through fine-tuning.

For simplicity, each task ends right after a click action is executed, and the agent can adjust the cursor position multiple times before making a click within the defined maximum steps quota.

We evaluate our approach on a custom benchmark of 430 single-click web tasks, each corresponding to a saved HTML page rendered in the VisionWebBrowser. For every page, we store the target element’s XPath -- a unique identifier within the DOM tree -- along with its bounding box and a short textual description automatically derived from visible text or metadata. Static HTML snapshots ensures that tasks remain stable over time, improving reproducibility. However, some dynamic artifacts (\eg, cookie banners or pop-ups) may still appear during rendering, occasionally causing occlusions or layout shifts. Since each target’s bounding box and XPath are preserved, we automatically detect when an element becomes obscured or replaced and exclude such tasks from evaluation.

In our experiments, for each task, we consider two configuration settings for the agent's input and prompting:
\begin{itemize}[leftmargin=*, topsep=2pt]
    \item \textbf{Interaction trace: hidden \vs. visible}. The agent either receives only the current screenshot and task description, or additionally the full interaction trace (intermediate reasoning and previously executed mouse actions).
    \item \textbf{Behavioral guidance: absent \vs. present}. We optionally include explicit instructions requiring the agent to reason whether the cursor is currently positioned over the target element and avoid issuing a combined move+click action in a single step.
\end{itemize}

\noindent We evaluate these settings under two \textbf{task formulations}:
\begin{itemize}[leftmargin=*, topsep=2pt]
    \item \textbf{Simplified}, automatically generated from the stored textual description: \textit{``Click on the element that displays \{text\} or conveys its meaning.''}
    \item \textbf{Human-like}, authored by Qwen2.5-VL-32B to emulate natural user queries, \eg, \textit{``Switch the website's language to English.''} for the target element displaying ``English''.
\end{itemize}

\noindent Each task is executed five times, and we report the average success rate. A click is considered successful if the cursor lands within the target bounding box. 

As shown in Table~\ref{tab:singleclick_simplified}, the base Qwen2.5-VL-32B agent reaches the success rate of 0.91--0.92 with interaction trace hidden.
When past thoughts and actions remain visible, its accuracy drops to 0.86--0.87, which confirms the earlier observation that the agent becomes distracted by its own reasoning.
Prompting the agent to reason about the cursor–target relation with interaction history present yields a marginal improvement by 1\%, indicating that handcrafted guidance provides only limited relief and that trace-driven bias is a key contributor to the performance drop.

%% file: fig/main_fig_tikz.tex
\definecolor{darkgreen}{HTML}{006400}
\tikzset{
    text box/.style={
        draw, thick, rounded corners=6pt,
        align=center, font=\sffamily\large, 
        inner xsep=0em},
    col1/.style={
        text width=4.3cm, minimum height=8mm},
    col2/.style={
        text width=2.8cm, minimum height=12mm},
    col3/.style={
        text width=2.3cm, minimum height=10mm},
    connector/.style={
        rounded corners=8pt, line width=1pt, 
        shorten >=1mm, shorten <=2mm,
        >={[width=7pt,length=9pt]Stealth}},
    annotate/.style={
        font=\sffamily, align=left, midway,
        inner xsep=1em, very thin},  
    subtitle/.style={
        anchor=base, inner ysep=0.5em,
        font=\sffamily\bfseries\large},
}
\begin{tikzpicture}
    \node[text box, col1] (QA) {Q\&A Task\\[-1mm]{\scriptsize Is the cursor pointing at the element?}\\[-2mm]{\scriptsize YES: click, NO: move}};

    \node[text box, col1, below=1mm of QA] (S1) {Screen};

    \node (GROUP1) [
        draw,
        rectangle,
        dashed,
        inner sep=4pt,
        fit=(QA)(S1)
        ] {};

    \node[text box, col1, inner ysep=.8em, below=30mm of S1] (SCT) {Single-Click Task\\[-1mm]{\scriptsize Click on the element}};

    \node[text box, col1, below=1mm of SCT] (S2) {Screen};

    \node (GROUP2) [
        draw,
        rectangle,
        dashed,
        inner sep=4pt,
        fit=(SCT)(S2)
        ] {};

    \node[text box, col1, below=1.5mm of GROUP2, text width=4.6cm, text=red, draw=red] (H) {Interaction Trace};

    \node[text box, col2, above=1.5mm of GROUP2, text=darkgreen, text width=4.6cm, draw=darkgreen] (Guidlines) {Guidelines\\[-1mm]{\scriptsize - reason about the cursor-target relation}\\[-2mm]{\scriptsize - do not combine move and click in one step}};

    \node[text box, col2, minimum height=10mm, anchor=north west, at=(GROUP1.north east), xshift=20mm] (T1) {Teacher};
    \node[text box, col2, minimum height=10mm, anchor=south west, at=(GROUP1.south east), xshift=20mm] (S1) {Student};
    \node[text box, col2, above=5mm of T1, text=darkgreen, draw=darkgreen] (Hints) {Hints\\[-1mm]{\scriptsize - target element \{x;y\}}\\[-2mm]{\scriptsize - correct action}};

    \node[text box, col2, minimum height=10mm, anchor=north west, at=(GROUP2.north east), xshift=20mm] (T2) {Teacher};
    \node[text box, col2, minimum height=10mm, anchor=south west, at=(GROUP2.south east), xshift=20mm] (S2) {Student};

    \coordinate (midloss1) at ($(T1)!0.5!(S1)+ (18mm,0)$);
    \node[text box, col3, right=18mm of T1] at (midloss1) (L1) {Loss};

    \coordinate (midloss2) at ($(T2)!0.5!(S2)+ (18mm,0)$);
    \node[text box, col3, right=18mm of T2] at (midloss2) (L2) {Loss};

    \draw[connector,->] (GROUP1.east) -- (T1.west);
    \draw[connector,->] (GROUP1.east) -- (S1.west);
    \draw[connector,->] (Hints.south) -- (T1.north);
    \draw[connector,->] (T1.east) -- (L1.west);
    \draw[connector,->] (S1.east) -- (L1.west);

    \draw[connector,->] (GROUP2.east) -- (T2.west);
    \draw[connector,->] (GROUP2.east) -- (S2.west);
    \draw[connector,->] (Guidlines.east) -- (T2.west);
    \draw[connector,->] (T2.east) -- (L2.west);
    \draw[connector,->] (S2.east) -- (L2.west);
    \draw[connector,->] (H.east) -- (S2.west);

    \path (Hints.north) coordinate[yshift=1mm] (TOP);
    \path (H.south) coordinate[yshift=-3mm] (BOTTOM);
    \path (L1.east) coordinate[xshift=-1mm] (RIGHT);
    \path (GROUP1.west) coordinate[xshift=-2mm] (LEFT);
    \path (S1.south) -- (Guidlines.north) coordinate[pos=0.5] (MID);
    \begin{scope}[on background layer]
        \draw[dashed,thick] (RIGHT |- MID) -- (LEFT |- MID);
    \end{scope}

    \node[subtitle,above=20mm of L1.north] (Stage1) {Stage 1};
    \node[subtitle,above=20mm of L2.north] (Stage2) {Stage 2};
    
\end{tikzpicture}

%% file: method.tex
\subsection{Task and Evaluation Setup}
\label{subsec:test_setup}

To improve interaction accuracy and mitigate the bias from prior actions and the high prompt sensitivity observed in preliminary experiments, we propose a fine-tuning procedure, illustrated in Figure~\ref{fig:diagram}, that strengthens spatial grounding and reasoning robustness.

Following~\cite{snell2022learning,kujanpaa2025efficient,alakuijala2025memento}, we adopt a self-distillation setup in which the teacher and student share the same architecture (Qwen2.5-VL-32B), but the teacher has access to privileged information, \textit{hints}.
Training minimizes the KL divergence between their output distributions, allowing the student to imitate the teacher without human supervision.

For both stages, we construct datasets using the VisionWebBrowser environment to ensure consistency between training and evaluation (see Section~\ref{subsec:test_setup}). The collected URLs cover diverse website layouts and content types.

We apply the same training setup across stages 1 and 2: LoRA~\cite{hu2022lora} adapters (rank 32) are added to the vision and language encoders, and models are trained with AdamW (weight decay $0.02$) using a 30-step linear warm-up to a peak learning rate of $10^{-4}$, followed by linear decay.
In line with~\cite{alakuijala2025memento}, we employ a high student dropout rate of 0.9 for effective self-distillation.

\subsection{Stage 1: Learning to Act from the Screen}
\label{subsec:stage1}
The first stage teaches the model to select the next action from the current visual state, deciding between click and move based on the cursor's position relative to the target element, and to predict target coordinates when movement is required.
Each training sample includes a task instruction identifying the target element, short behavioral guidance aligned with the VisionWebBrowser interaction loop (see Section~\ref{sec:environment}), and a screenshot with the cursor positioned at varying distances from the target. The teacher receives privileged hints containing the correct action and the ground-truth cursor and target coordinates, while the student observes only the screenshot and the task description.
This stage can be viewed as a question–answering training that strengthens visual grounding and action control.

\begin{table}[tbp]
\caption{Accuracy on the Stage~1 move-or-click selection task.}
\label{tab:qa_step1}
\centering
\setlength{\tabcolsep}{9.4pt}
\begin{tabular}{cccc}
\toprule
& \textbf{Baseline} & \textbf{Stage 1} & \textbf{Teacher} \\
\midrule
move \vs. click & 0.79 & 0.94 & 0.99 \\
\bottomrule
\end{tabular}

\parbox{\linewidth}{\footnotesize \textit{Baseline} = original Qwen2.5-VL-32B; \textit{Stage~1} = fine-tuned for move-or-click action selection; \textit{Teacher} = baseline with access to the correct action and target coordinates.}
\end{table}

\begin{table}[tbp]
\caption{Accuracy on single-click tasks with \textit{simplified / human-like task descriptions}.}
\label{tab:singleclick_simplified}
\centering
\setlength{\tabcolsep}{3.2pt}

\begin{tabular}{ccccc}
\toprule
 \textbf{Interaction} & \textbf{Behavioral} & \multirow{2}{*}{\textbf{Baseline}} & \multirow{ 2}{*}{\textbf{Stage 1}} & \multirow{2}{*}{\textbf{Stage 2}} \\
 \textbf{Trace} & \textbf{Guidance} & & & \\
\midrule
\ding{55} & \ding{52} & 0.92 / 0.91 & \textbf{0.94 / 0.94} & -- \\
\ding{52} & \ding{52} & 0.87 / 0.87 & 0.93 / 0.91 & \textbf{0.94 / 0.94} \\
\ding{52} & \ding{55} & 0.87 / 0.86\textsuperscript{1} & 0.92 / 0.91 & \textbf{0.95 / 0.94} \\
\bottomrule
\end{tabular}

\parbox{\linewidth}{\footnotesize
\textit{Interaction Trace} = past reasoning and actions; \textit{Behavioral Guidance} = instructions to check cursor–target alignment and avoid combining move and click; \textit{Baseline} = original Qwen2.5-VL-32B; \textit{Stage~1} = fine-tuned for move-or-click action selection; \textit{Stage~2} = trained for robustness to interaction-trace-induced bias and reduced prompting.
\textbf{Bold} indicates \textbf{the best performance} within each row. \textbf{95\% confidence interval half-widths are 0.01 for all estimates, except for the one marked with \textsuperscript{1}, where the half-width is 0.02.}}
\end{table}

\paragraph{Experimental Results.} Before applying the model to the single-click benchmark, we first evaluate it on a validation set from the Stage~1 question-answering dataset, allowing us to assess action selection accuracy outside the agentic browser environment. As shown in Table~\ref{tab:qa_step1}, the base Qwen2.5-VL-32B model reaches 0.79 accuracy on the move-or-click action selection task, while the Stage~1 tuned model improves to 0.94, substantially reducing the gap to the teacher’s score of 0.99. This confirms that Stage~1 strengthens the model’s ability to read the screen and choose the correct action. 
Since single-click task success requires an accurate mouse move followed by a click, action-selection accuracy provides a natural upper bound on downstream performance ($\approx 0.94$).

We next evaluate the model on the full single-click benchmark (Table~\ref{tab:singleclick_simplified}). With the interaction trace hidden, Stage~1 fine-tuning increases success rates from 0.86--0.87 to 0.94, reaching the expected upper limit. When the trace is visible, the success rate of the baseline model drops by about 5 percentage points, reflecting the distraction caused by previous reasoning traces. The Stage~1 model, however, shows reduced sensitivity to such noise and achieves 0.91--0.93, suggesting that more reliable move-or-click decisions enhance the model's focus on the current screen.

Nevertheless, even with explicit behavioral guidance instructing the model to analyze the cursor–target relation and separate movement from clicking, the presence of interaction history continues to degrade performance (row 2 of Table~\ref{tab:singleclick_simplified}).

\subsection{Stage 2: Focusing on the Screen State}
\label{subsec:stage2}

\begin{table}[tbp]
\caption{Closed-loop correction rate with 95\% Clopper-Pearson confidence intervals.}
\label{tab:closed_loop}
\begin{tabular}{cccc}
\toprule
& \multicolumn{1}{c}{\textbf{Baseline}} 
& \multicolumn{1}{c}{\textbf{Stage 1}} 
& \multicolumn{1}{c}{\textbf{Stage 2}} \\
\midrule
$R_{\text{corr}}$ &
1.9\% \ {\scriptsize 0.7\%--4.3\%} &
0.0\% \ {\scriptsize 0.0\%--2.0\%} &
35.0\% \ {\scriptsize 28.1\%--42.4\%} \\
\bottomrule
\end{tabular}

\parbox{\linewidth}{\footnotesize\textit{Baseline} = original Qwen2.5-VL-32B; \textit{Stage~1} = fine-tuned for move-or-click action selection; \textit{Stage~2} = trained for robustness to history-induced bias and reduced prompting. Interaction Trace: \ding{52}, Behavioral guidance: \ding{52}, human-like task formulations.}
\end{table}

The second stage addresses the model’s tendency to follow initial reasoning despite webpage changes and its dependence on handcrafted reasoning and behavior prompts.
To collect teacher demonstrations, we hide the interaction trace for the Stage~1 agent and ask it to reason about the cursor position relative to the element of interest before acting while completing the training set of single-click tasks in our VisionWebBrowser environment.
We then reveal that trace to the student, so that it now has access to the ``privileged'' bias-inducing information, and remove the detailed thinking instructions from the student’s prompt, leaving only minimal formatting guidance.
Thus, the teacher generates unbiased outputs following the reasoning guidelines, and the student learns to reproduce them despite additional contextual trace-induced noise and reduced prompting, which enhances focus and robustness.

\paragraph{Experimental Results.} 
As shown in Table~\ref{tab:singleclick_simplified}, Stage 2 training, specifically designed to mitigate the bias toward previous interaction traces and prompt dependence, achieves success rates of 0.94--0.95 in the most challenging setting, outperforming the baseline and Stage~1 models by 8\% and 3\%, respectively.

Beyond success rates, we further analyze how Stage~2 changes agent interaction patterns.
Since Stage~2 trains the model to internalize guidelines for verifying whether the cursor is currently pointing at the target element before acting, we expect it to develop a closed-loop control habit of reflecting on the current state and readjusting the cursor when necessary. To assess whether the Stage~2 model indeed learns this behavior, we compute the \textbf{closed-loop correction rate} $R_{\text{corr}}$ as a percentage of tasks in which the first mouse move is inaccurate (outside the bounding box of the target element), but the agent adjusts the coordinates in the next steps and completes the task successfully.
As summarized in Table~\ref{tab:closed_loop}, the baseline and Stage~1 models almost never correct the false prediction of the initial move coordinates ($R_{\text{corr}} \approx 0\%$).
Further inspection suggests that they rarely revise a move and proceed directly to clicking, so their success is largely determined by move accuracy, which explains why the Stage~1 performance reaches its upper bound.
In contrast, Stage~2 achieves $R_{\text{corr}} = 35\%$, indicating that it partially adopts the closed-loop behavior pattern.

Finally, across all configurations and models, the difference between simplified and human-like task descriptions remains within 1--2\%.
Although small, this gap is consistent and suggests that linguistic variability may still affect task understanding.

%% file: conclusion.tex
Our findings confirm the feasibility of vision-only single-module agents. The proposed two-stage self-distillation improves spatial grounding and reduces reliance on external guidance and previous reasoning and actions, leading to consistently better performance on our benchmark.
Beyond accuracy gains, our training partially enables closed-loop behavior, where the agent adjusts its actions based on the current screen.

While limited to single-click interactions, the framework provides a meaningful basis for scaling to broader action spaces and multi-step tasks. 
Future work will further develop closed-loop control and extend the approach to more dynamic settings.

%% file: references.bib
@misc{koh2024visualwebarena,
  title={{VisualWebArena}: Evaluating Multimodal Agents on Realistic Visual Web Tasks},
  author={Koh, Jing Yu and Lo, Robert and Jang, Lawrence and Duvvur, Vikram and Lim, Ming Chong and Huang, Po-Yu and Neubig, Graham and Zhou, Shuyan and Salakhutdinov, Ruslan and Fried, Daniel},
  archivePrefix={arXiv},
  primaryClass={cs.CL},
  eprint={2401.13649},
  year={2024},
  url={https://arxiv.org/abs/2401.13649},
}

@inproceedings{zhou2024webarena,
  title={{WebArena}: A Realistic Web Environment for Building Autonomous Agents},
  author={Zhou, Shuyan and Xu, Frank F and Zhu, Hao and Zhou, Xuhui and Lo, Robert and Sridhar, Abishek and Cheng, Xianyi and Bisk, Yonatan and Fried, Daniel and Alon, Uri and others},
  booktitle={Proceedings of the International Conference on Learning Representations},
  year={2024},
  url={https://webarena.dev},
}

@misc{deng2023mind2web,
  title={{Mind2Web}: Towards a Generalist Agent for the Web},
  author={Xiang Deng and Yu Gu and Boyuan Zheng and Shijie Chen and Samuel Stevens and Boshi Wang and Huan Sun and Yu Su},
  year={2023},
  eprint={2306.06070},
  archivePrefix={arXiv},
  primaryClass={cs.CL},
  url={https://arxiv.org/abs/2306.06070},
}

@misc{he2024webvoyager,
  title={{WebVoyager}: Building an End-to-End Web Agent with Large Multimodal Models},
  author={He, Hongliang and Yao, Wenlin and Ma, Kaixin and Yu, Wenhao and Dai, Yong and Zhang, Hongming and Lan, Zhenzhong and Yu, Dong},
  year={2024},
  eprint={2401.13919},
  archivePrefix={arXiv},
  primaryClass={cs.CL},
  url={https://arxiv.org/abs/2401.13919},
}

@misc{niu2024screenagent,
  title={{ScreenAgent}: A Vision Language Model-driven Computer Control Agent}, 
  author={Runliang Niu and Jindong Li and Shiqi Wang and Yali Fu and Xiyu Hu and Xueyuan Leng and He Kong and Yi Chang and Qi Wang},
  year={2024},
  eprint={2402.07945},
  archivePrefix={arXiv},
  primaryClass={cs.HC},
  url={https://arxiv.org/abs/2402.07945}
}

@misc{bhathal2025websight,
  title={{Websight}: A vision-first architecture for robust web agents},
  author={Bhathal, Tanvir and Gupta, Asanshay},
  year={2025},
  eprint={2508.16987},
  archivePrefix={arXiv},
  primaryClass={cs.AI},
  url={https://arxiv.org/abs/2508.16987}, 
}

@inproceedings{zheng2024seeact,
  title={{GPT}-4V(ision) is a Generalist Web Agent, if Grounded},
  author={Boyuan Zheng and Boyu Gou and Jihyung Kil and Huan Sun and Yu Su},
  booktitle={Forty-first International Conference on Machine Learning},
  year={2024},
  url={https://openreview.net/forum?id=piecKJ2DlB},
}

@misc{qwen2.5-VL,
    title = {Qwen2.5-VL},
    url = {https://qwenlm.github.io/blog/qwen2.5-vl/},
    author = {Qwen Team},
    month = {jan},
    year = {2025},
}

@inproceedings{yao2023react,
  title = {{ReAct}: Synergizing Reasoning and Acting in Language Models},
  author = {Yao, Shunyu and Zhao, Jeffrey and Yu, Dian and Du, Nan and Shafran, Izhak and Narasimhan, Karthik and Cao, Yuan},
  booktitle = {Proceedings of the International Conference on Learning Representations},
  year = {2023},
  html = {https://arxiv.org/abs/2210.03629},
}

@article{kujanpaa2025efficient,
    title={Efficient Knowledge Injection in {LLM}s via Self-Distillation},
    author={Kalle Kujanp{\"a}{\"a} and Pekka Marttinen and Harri Valpola and Alexander Ilin},
    journal={Transactions on Machine Learning Research},
    issn={2835-8856},
    year={2025},
    url={https://openreview.net/forum?id=drYpdSnRJk},
}

@misc{alakuijala2025memento,
      title={Memento No More: Coaching AI Agents to Master Multiple Tasks via Hints Internalization}, 
      author={Minttu Alakuijala and Ya Gao and Georgy Ananov and Samuel Kaski and Pekka Marttinen and Alexander Ilin and Harri Valpola},
      year={2025},
      eprint={2502.01562},
      archivePrefix={arXiv},
      primaryClass={cs.LG},
      url={https://arxiv.org/abs/2502.01562}, 
}

@inproceedings{hu2022lora,
    title={Lo{RA}: Low-Rank Adaptation of Large Language Models},
    author={Edward J Hu and Yelong Shen and Phillip Wallis and Zeyuan Allen-Zhu and Yuanzhi Li and Shean Wang and Lu Wang and Weizhu Chen},
    booktitle={Proceedings of the International Conference on Learning Representations},
    year={2022},
    url={https://openreview.net/forum?id=nZeVKeeFYf9}
}

@misc{snell2022learning,
      title={Learning by Distilling Context}, 
      author={Charlie Snell and Dan Klein and Ruiqi Zhong},
      year={2022},
      eprint={2209.15189},
      archivePrefix={arXiv},
      primaryClass={cs.CL},
      url={https://arxiv.org/abs/2209.15189}, 
}

@inproceedings{cheng2024seeclick,
    title = "{S}ee{C}lick: Harnessing {GUI} Grounding for Advanced Visual {GUI} Agents",
    author = "Cheng, Kanzhi  and
      Sun, Qiushi  and
      Chu, Yougang  and
      Xu, Fangzhi  and
      YanTao, Li  and
      Zhang, Jianbing  and
      Wu, Zhiyong",
    booktitle = "Proceedings of the 62nd Annual Meeting of the Association for Computational Linguistics",
    month = aug,
    year = "2024",
    address = "Bangkok, Thailand",
    publisher = "Association for Computational Linguistics",
    url = "https://aclanthology.org/2024.acl-long.505",
    pages = "9313--9332"
}

@misc{qin2025ui,
      title={UI-TARS: Pioneering Automated GUI Interaction with Native Agents}, 
      author={Qin, Yujia and Ye, Yining and Fang, Junjie and Wang, Haoming and Liang, Shihao and Tian, Shizuo and Zhang, Junda and Li, Jiahao and Li, Yunxin and Huang, Shijue and others},
      year={2025},
      eprint={2501.12326},
      archivePrefix={arXiv},
      primaryClass={cs.AI},
      url={https://arxiv.org/abs/2501.12326}, 
}
